\documentclass{elsart}
\journal{Astroparticle Physics}
\usepackage{epsfig}
\usepackage{multirow}

\DeclareSymbolFont{AMSa}{U}{msa}{m}{n}
\DeclareMathSymbol{\lesssim}{\mathrel}{AMSa}{"2E}
\DeclareSymbolFont{AMSb}{U}{msb}{m}{n}
\DeclareMathSymbol{\varnothing}{\mathord}{AMSb}{"3F}

\newcommand\degree{\ensuremath{^\circ}}
\newcommand\diameter{\ensuremath{\varnothing}}

\begin{document}
\begin{frontmatter}
\title{Using the photons from the Crab Nebula seen by GLAST to
       calibrate MAGIC and the Imaging Air Cherenkov Telescopes}
\author[PD]{D.~Bastieri\corauthref{cor}},
\corauth[cor]{Corresponding author.}
\ead{bastieri@pd.infn.it}
\author[PD]{C.~Bigongiari},
\author[UD]{E.~Bisesi},
\author[PD]{G.~Busetto},
\author[UD]{A.~De~Angelis},
\author[UD]{B.~De~Lotto},
\author[UD]{T.~Lenisa},
\author[TS]{F.~Longo},
\author[PD]{M.~Mariotti},
\author[PD]{A.~Moralejo},
\author[PD]{D.~Pascoli},
\author[PD]{L.~Peruzzo},
\author[UD]{S.~Raducci},
\author[PD]{R.~Rando},
\author[PD]{A.~Saggion},
\author[PD]{P.~Sartori} and
\author[PD]{V.~Scalzotto}.
\address[PD]{Universit\`a di Padova and INFN Padova, Italy.}
\address[TS]{Universit\`a di Trieste and INFN Trieste, Italy.}
\address[UD]{Universit\`a di Udine and INFN Trieste, Italy.}
\begin{abstract}
In this article we discuss the possibility of using the observations by
GLAST of standard gamma sources, as the Crab Nebula, to calibrate
Imaging Air Cherenkov detectors, MAGIC in particular, and optimise
their energy resolution. We show that at around $100\:\mathrm{GeV}$ the
absolute energy calibration uncertainty of Cherenkov telescopes can be
reduced to $\lesssim 10\%$ by means of such cross-calibration
procedure.
\end{abstract}
\end{frontmatter}

\section{Introduction}\label{s:intro}
Full multiwavelength coverage over as wide an energy range as possible
is needed to understand aspects of fundamental physics and astrophysics
as well.  An important observational window, between few tens and some
hundreds GeV, is still largely unknown due to experimental detection
difficulties, related to the opacity of our atmosphere to gamma-rays.
For this reason, observations have to be performed:
\begin{itemize}
\item on board of satellites orbiting outside our atmosphere (given the
 steep decrease of the gamma-ray flux, the limited size of the
 detectors, due to launching cost, sets an upper limit to the
 accessible energy region),
\item detecting, on the ground, showers initiated by gamma-rays in the
 atmosphere (in this case, as the measurement of the gamma-ray
 properties are done exploiting the atmosphere as a calorimeter, there
 is a lower energy limit to the observable gamma-rays).
\end{itemize}
\begin{figure}[tbp]
\center{\epsfig{file=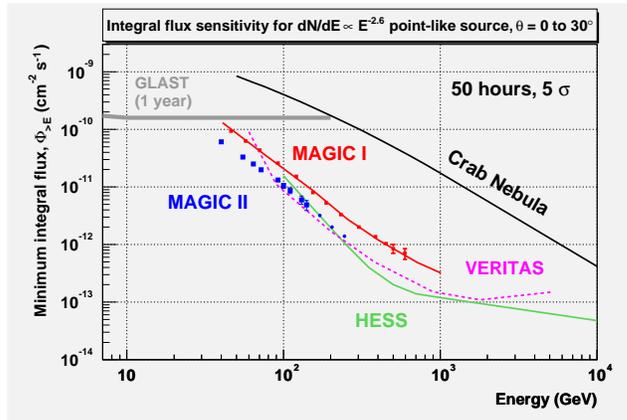,width=0.6\textwidth}}
\caption{\label{fig:sensi}Predicted sensitivities for some operating
 and proposed detectors. Note the wide overlap region between GLAST
 and present Cherenkov telescopes. As far as MAGIC is concerned, the
 solid, red line represents the predictions made by the full
 Montecarlo simulation and it is in good agreement with the
 sensitivity calculated from the first observations.  The blue dots
 are the expected sensitivity for MAGIC II, where a second telescope,
 \emph{clone\/} of the current MAGIC, will be built at
 $\sim 85\:\mathrm{m}$ of distance from MAGIC.  Start of operation
 for MAGIC II is envisaged for the end of 2006, just before the
 scheduled launch of GLAST.}
\end{figure}

Among ground-based detectors, the so-called IACTs (Imaging Atmospheric
Cherenkov Telescopes) are expected to reach the lowest energy
thresholds.  IACTs feature huge collection areas, an excellent angular
resolution and good energy confinement even at very high energies.  On
the other hand, they suffer from low duty-cycles, small fields of view
($<$5\degree) and systematic calibration uncertainties in both
energy and sensitivity. In fact, whereas current IACTs achieve an
intrinsic energy resolution as low as 5\%, the absolute energy scale
remains quite elusive, as the energy reconstruction in the
$30\div 300\:\mathrm{GeV}$ range is dominated by uncertainties on
Monte Carlo simulations and on atmospheric models\cite{moh}.

In the next few years, two satellites with a payload dedicated to
observation of gamma rays will be launched: AGILE\cite{agile}, a small
instrument expected to fly in 2005, and GLAST\cite{glast}, a large
detector whose launch is scheduled for the beginning of the year 2007.
GLAST will operate in a complementary manner to ground-based
experiments and provide ground-based observers with alerts
for transient sources, as new imaging Cherenkov telescopes are built
to slew toward a source within few tens of seconds following a prompt
notification.

\begin{figure}
\center{\epsfig{file=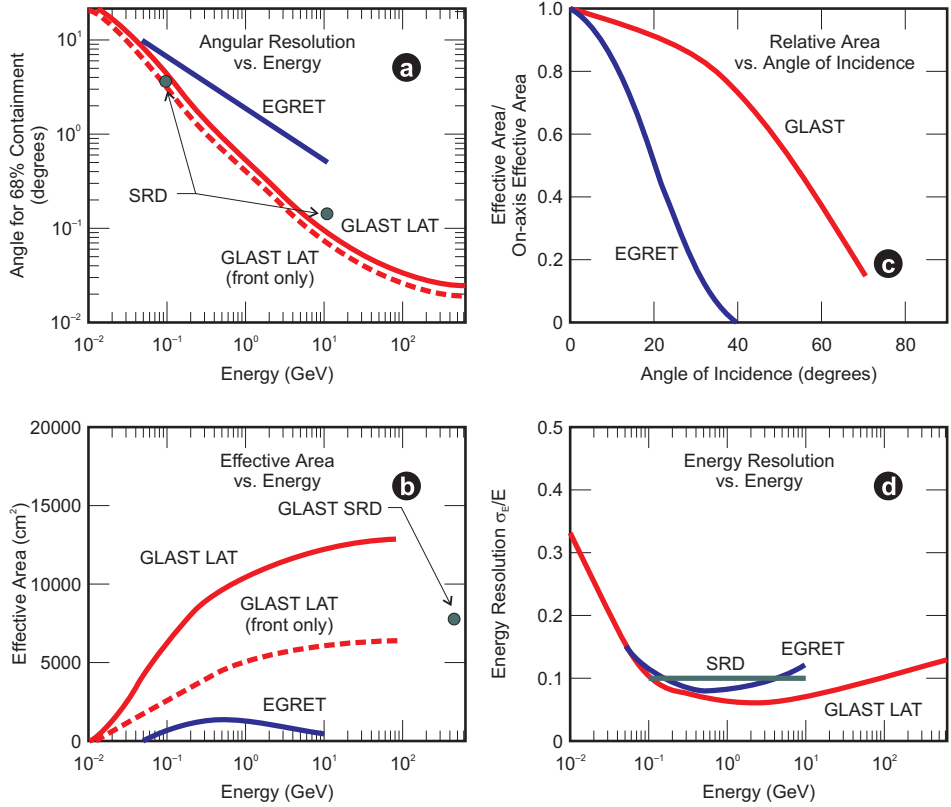,width=0.9\textwidth}}
\caption{\label{fig:glast}Full \emph{Instrument Response Functions\/}
 (IRFs) of the GLAST LAT (taken from
 \emph{http://www-glast.slac.stanford.edu/software/IS/glast%
       \_lat\_performance.htm}).}
\end{figure}

Satellites as AGILE or GLAST, contrary to IACTs, are calibrated in a
well-controlled laboratory environment, using test beams of electrons
and gammas, and a relative uncertainty of $\sim 10\%$ or better is
expected (see Figure~\ref{fig:glast}d).  After GLAST launch, while
LIDARs can provide IACTs with regular measurements of atmospheric
transmission, GLAST observations of steady sources at the highest
energies can be used to reduce systematic errors in the absolute
energy scale determination of IACT events.

Four major IACTs for the observation of gamma rays, designed as arrays
of 10-meter class mirrors with finely pixelized imaging cameras, just
started running: MAGIC\cite{magic} and VERITAS\cite{veritas}in the
northern hemisphere; HESS\cite{hess} and CANGAROO-III\cite{cangaroo} in
the southern one. In particular MAGIC, which started operations in
October 2003 at La Palma, Canary Islands, consists of a single
$17\:\mathrm{m}\:\diameter$ tessellated mirror.  As the largest
instrument, it can access the lowest threshold, and has the widest
overlap with satellites (see Figure~\ref{fig:sensi} from \cite{sensi}).
Its actual energy threshold is expected to be, at regime, as low as
$30\:\mathrm{GeV}$, although an even lower threshold of few GeV is not
excluded, provided a new high quantum efficiency camera is implemented.
Standing the present panorama, MAGIC can benefit most from GLAST data.

In this paper, we discuss the possibility of using GLAST observations
of a standard gamma source, the Crab Nebula, to calibrate the Imaging
Air Cherenkov detectors, MAGIC in particular, in order to extract the
maximal information from the GeV region.

The paper is organised as follows.  After this introduction, in Section
\ref{s:glast} we estimate the number of photons from Crab Nebula seen
by GLAST.  In Section \ref{s:cross} we outline a possible strategy for
cross-calibration based on the observations by GLAST. Finally, in
Section \ref{s:conc} we draw our conclusions.

\section{Detection of Crab Nebula by GLAST}\label{s:glast}
The performance of GLAST was studied by means of a full simulation
based on Geant4\cite{perugia}.  The on-axis effective area of GLAST as
a function of energy is shown in Figure~\ref{fig:glast}b (from
Ref.~\cite{glast}), with an average value for the GLAST Large Area
Telescope (LAT) of about $1.3\:\mathrm{m}^2$ around
$100\:\mathrm{GeV}$.
\begin{figure}[tbp]
\center{\epsfig{file=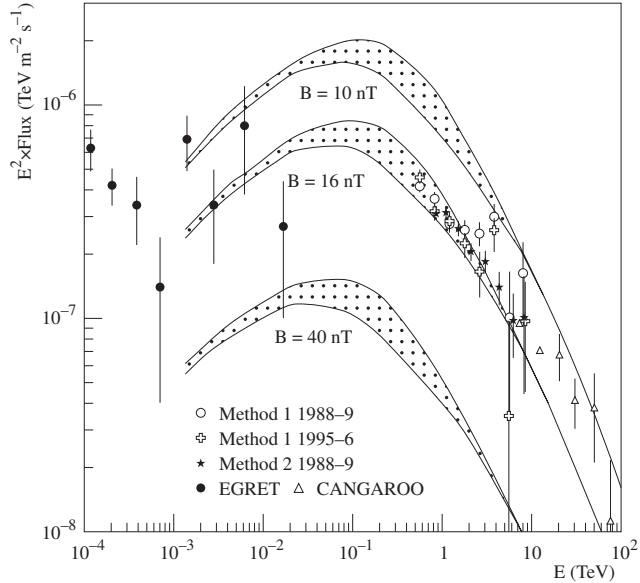,width=0.6\textwidth}}
\caption{\label{crab}Expected gamma spectrum from Crab Nebula.}
\end{figure}

During the first year, GLAST will observe the sky in survey (scanning)
mode, therefore a uniform exposure at a 90\% level can be
conservatively assumed\cite{glastsciencedoc}.  As its field of view is
around $2.4\:\mathrm{sr}$, \emph{i.e.}, $\sim\frac{1}{5}$ of the full
sky, GLAST will observe every source, and in particular the Crab
Nebula, for $\frac{1}{5}$ of a year. Most of the time the source will
be off-axis by $40\degree$ on average, and the effective area is
correspondingly reduced by a factor of 0.8 as seen in
Figure~\ref{fig:glast}c.

The spectrum of the Crab Nebula in the overlap region is poorly known:
under different hypotheses on the magnetic field in an Inverse Compton
scenario, it changes according with Figure~\ref{crab} (from
Ref.~\cite{hillas}).  The variation in spectral index, from lower to
higher energies, can be used to define a unique energy scale. In fact,
the spectrum can be parameterised with two different spectral indexes:
one fitting data at low energies and one at higher energies.  We can
define $E_{\mathrm{brk}}$ as the energy at which the two power laws
meet.  Let us assume, conservatively, that the low energy spectral
index is 2.0 and the high energy one is 2.7.  A bigger difference
between the indexes will mark even more the spectral feature and make
the determination of $E_{\mathrm{brk}}$ easier. The value of
$E_{\mathrm{brk}}$ is of the order of magnitude of
$100\:\mathrm{GeV}$.  \emph{The position of this spectral break, well
determined by GLAST, can be used to calibrate MAGIC.}

The number of photons from Crab Nebula between 10 and
$300\:\mathrm{GeV}$ detected in the first year by GLAST in survey mode
(with a 90\% data efficiency allowing for South Atlantic Anomaly
passages, data downlink failures etc.), depends on $E_{\mathrm{brk}}$.
The actual value obtained from the simulation, as a function of
$E_{\mathrm{brk}}$, is listed in Table~\ref{tab:flux}.
\begin{table}
\renewcommand{\multirowsetup}{\centering}
\caption{\label{tab:flux}Number of photons from Crab Nebula detected by
         GLAST in one year and relative error on the determination of
         $E_{\mathrm{brk}}$.  MAGIC is assumed to collect 50,000 gammas
         in 50 hours and the error on $E_{\mathrm{brk}}$ takes into
         account only the statistics as explained in the text.}
\begin{tabular}{c|c|p{2cm}|p{2cm}c}
\cline{1-4}
  \multirow{2}{2cm}{$E_{\mathrm{brk}}$ (GeV)}
  & \multirow{2}{2cm}{Gammas seen by GLAST}
  & \multicolumn{2}{c}{$\delta E_{\mathrm{brk}}/E_{\mathrm{brk}}$} \\
\cline{3-4}
  && \centering GLAST & \centering MAGIC & \\
\cline{1-4}
 ~50 & 3763 & \centering ~6.2\% & \centering 4.0\% & \\
 100 & 3249 & \centering ~8.2\% & \centering 3.5\% &\\
 150 & 2988 & \centering 12.7\% & \centering 2.9\% &\\
 200 & 2818 & \centering 17.2\% & \centering 5.2\% &\\
\cline{1-4}
\end{tabular}
\end{table}

\section{Calibration Strategy for IACT}\label{s:cross}
We simulated the observation by GLAST in the range from 10 to
$300\:\mathrm{GeV}$ and assumed the energy resolution in
Figure~\ref{fig:glast}d.  The photon samples followed the spectra
expected from Crab Nebula at different values of $E_{\mathrm{brk}}$ and
with the multiplicities of Table~\ref{tab:flux}.

The value of $E_{\mathrm{brk}}$ was then estimated from each sample
by minimising the residual values, properly weighted, of the simulated
data from some \emph{template\/} distributions that mimicked the Crab
spectrum at different values of $E_{\mathrm{brk}}$.  The error on the
determination of $E_{\mathrm{brk}}$ was calculated from the spread of
the central values of the estimate obtained from 100 independent Monte
Carlo samples. The relative errors on $E_{\mathrm{brk}}$ obtained with
such a procedure are listed in the column headed GLAST of
Table~\ref{tab:flux}.

On the other side, we simulated for MAGIC a total flux of 50,000 gamma
photons coming from the Crab Nebula in 50 hours of observation time.
The estimate of 1000 gammas/hour should be easily attainable at regime.
To unfold the simulated distribution we filled a migration matrix using
the full Montecarlo simulation and then applied it to the simulated
spectrum.  Applying the same algorithm for the determination of
$E_{\mathrm{brk}}$ we obtain the relative error for MAGIC as listed in
the last column of Table~\ref{tab:flux}.

This error is taking into account only the statistics, and it is quite
lower than the GLAST one.

The spectral feature $E_{\mathrm{brk}}$, as seen by GLAST, can be used
to calibrate MAGIC: the value of $E_{\mathrm{brk}}$ as determined by
MAGIC should be offset to match GLAST value.  In this way the absolute
scale uncertainty in the region between 30 and $200\:\mathrm{GeV}$ will
not exceed GLAST one, \emph{i.e.}, it will go from about 6\% to 17\%.
Higher values of $E_{\mathrm{brk}}$ will be hardly reconstructed by
GLAST due to the steep decrease in the power-law spectrum, as can be
inferred from the rapidly increasing values of
$\delta E_{\mathrm{brk}}/E_{\mathrm{brk}}$ as a function of
$E_{\mathrm{brk}}$.

\section{Conclusions}\label{s:conc}
We showed in this paper that we can use GLAST observations of a
standard gamma source, as the Crab Nebula, to calibrate the absolute
energy scale of Imaging Air Cherenkov detectors and MAGIC in 
particular. In this way, the absolute energy uncertainty of Cherenkov
telescopes, at around $100\:\mathrm{GeV}$, can be reduced to
$\lesssim 10\%$.  A spectral break at higher energies will be harder to
measure and of little help to IACTs.  Nevertheless, other features, as
the exponential cutoff of AGN spectra, due to the interaction of AGN
gammas with the extragalactic infrared background, can also be
well-suited for IACT calibration, provided they can be observed by
GLAST and IACTs in the same energy range.

\end{document}